\documentclass[useAMS,usenatbib,usegraphicx]{mn2e}%

\title[Periodic flaring rate on YY Gem]{Simulations of the periodic flaring rate on YY Gem}
\author[D. H. Gao, P. F. Chen, M. D. Ding, and X. D. Li]
{D. H. Gao\thanks{E-mail: mg0526001@smail.nju.edu.cn},
P. F.  Chen\thanks{chenpf@nju.edu.cn},
M. D. Ding\thanks{dmd@nju.edu.cn},
X. D. Li\\
Department of Astronomy, Nanjing University, Nanjing 210093,
China}

\begin{document}

\date{Accepted --. Received --; in original form --}

\pagerange{\pageref{firstpage}--\pageref{lastpage}} \pubyear{2007}

\maketitle

\label{firstpage}

\begin{abstract}
The binary YY Gem shows many interesting properties, one of which
is the periodicity in its flaring rate. The period, which is about
$48 \pm 3$ min, was ever interpreted in terms of the oscillation
of a filament. In this paper, we propose a new model to explain
this phenomenon by means of 2.5-dimensional MHD numerical
simulations. It is found that magnetic reconnection is induced as
the coronal loops rooted on both stars inflate and approach each
other, which is driven by the differential stellar rotation. The
magnetic reconnection is modulated by fast-mode magnetoacoustic
waves which are trapped between the surfaces of the two stars, so
that the reconnection rate presents a periodic behaviour. With the
typical parameters for the binary system, the observed period can
be reproduced. We also derive an empirical formula to relate the
period of the flaring rate to the coronal temperature and density,
as well as the magnetic field.
\end{abstract}

\begin{keywords}
binaries: close -- stars: flare -- stars: oscillations -- waves --
methods: numerical
\end{keywords}

\section{Introduction}

The binary system YY Gem has been studied for many years since discovered
\citep{Adams20}. It is a double-lined spectroscopic eclipsing binary
system which contains two dM1e late-type stars with masses and radii
almost identical, and the orbital period being $0.8142822$ d
(\citealt{Joy26}; \citealt{vanGent31}). It is of great importance to
study the basic relations of such rare
late-type binaries (\citealt{Kron52}; \citealt*{Butler96}).  Based
on some considerations in \citet{Dworak75}, \citet{Brancewicz80}
gave the parameters of YY Gem: $M_1 = M_2 = 0.57$ M$_{\odot}$,
$R_1 = R_2 = 0.6$ R$_{\odot}$, the separation of the two components $a
= 3.83$ R$_{\odot}$.

\citet{Chabrier95} demonstrated that the depth of each stellar
radiative core is $\sim$ 70 per cent of its radius, and the
thickness of the convective zone is $\sim$ 30 per cent. Strong
convective motions under the surfaces cause large-scale star spots
and huge magnetic field structures. Flare activity on YY Gem was
reported for the first time by \citet{Moffett71}. Subsequent
investigations have shown that it is one of the most active
flaring binaries (\citealt{Moffett74}; \citealt{Doyle85};
\citealt{Doyle90a}). Their flaring activities have been studied in
multi-wavelengths. Using the Very Large Array, \citet*{Jackson89}
presented radio observations; \citet{Stelzer02} studied the
simultaneous X-ray spectroscopy of YY Gem with \emph{Chandra} and
\emph{XMM-Newton} satellites; While, far UV observations of YY Gem
were reported by \citet{Saar03}. The flarings on YY Gem exhibit UV
and X-ray emissions which are stronger than those on the Sun
\citep{Haisch90}. For example, \citet{Tsikoudi00} observed two
large flares, and the integrated X-ray luminosity was estimated to
be about 6-8 $\times10^{33}$ erg. They also calculated the ratio
of X-ray and bolometric luminosities $L_{X} / L_{Bol}$, which
indicated strong magnetic activities and `hot' coronal components.

For close binaries with similar masses, the density scale-height
is quite large, i.e., the density does not decrease with height
from the stellar surfaces as rapidly as in the solar corona.
Therefore, the hot plasma between the two stars could emit strong
X-ray light. For example, observations revealed large-scale strong
X-ray sources between the two stars of the RS CVn binary AR Lac
(\citealt{Siarkowski92}; \citealt{Siarkowski96}), as well as of
the RS CVn binary TY Pyx (\citealt{Culhane90}; \citealt*{Pres95}).
At the same time, it is probable that the two stars are
magnetically connected, as illustrated by \citet{Uchida83}.
\citet{Uchida85} proposed that the interstellar activities are
intimately affected by the differential rotations of both stars.
So, there may exist some interesting phenomena in such a system.
One of them is the periodic flaring rate, which was reported by
\citet{Doyle90a}.  Their observation in the \emph{U}-band showed
four flares separated in succession with a periodicity of $48 \pm
3$ min during a total observing time of 408 min. The duration of
each flare varies from $\sim 20$ min to $\sim 40$ min, and the
time-averaged flare luminosity in the \emph{U}-band is 1.25
$\times$ $10^{28}$ erg s$^{-1}$. They interpreted this periodicity
in terms of filament oscillations. In this paper, we attempt to
propose an alternative explanation for the periodicity.

Flares on the Sun and dKe-dMe stars are both thought to result
from magnetic reconnection. On the Sun, periodic behaviours have
also been reported in various reconnection-associated phenomena,
e.g., the quasi-periodic modulation of flaring emission
(\citealt{Nakariakov06}) and the repetitive appearance of
transition region explosive events (\citealt{Chenpf06}). In the
former case, magnetic reconnection is modulated by magnetoacoustic
waves from a nearby oscillating coronal loop, whereas in the
latter, magnetic reconnection is modulated by {\it p}-mode waves.

To our knowledge, no efforts in MHD numerical simulations have
been devoted to the investigation of flaring phenomenon in binary
systems, although hydrodynamic simulations have been done
regarding flares in M dwarf stars (e.g. \citealt{Chengcc91}).
Using potential field extrapolation, \citet{Uchida85} demonstrated
that the evolving surface magnetic field can result in magnetic
reconnection in the interstellar space. Here, we perform MHD
numerical simulations in order to investigate the magnetic
reconnection in a close binary system, with the purpose to clarify
which process is likely to be responsible for the periodicity of
the flaring rate as reported by \citet{Doyle90a}. The paper is
organised as follows. Section 2 describes the basic model, MHD
equations, initial and boundary conditions. Section 3 gives the
numerically calculated results. Discussions are presented in
section 4, which is followed by conclusions in section 5.

\section[]{numerical method}
\subsection{Problem setup}

\begin{figure*}
  \centering
  \includegraphics[width=14cm]{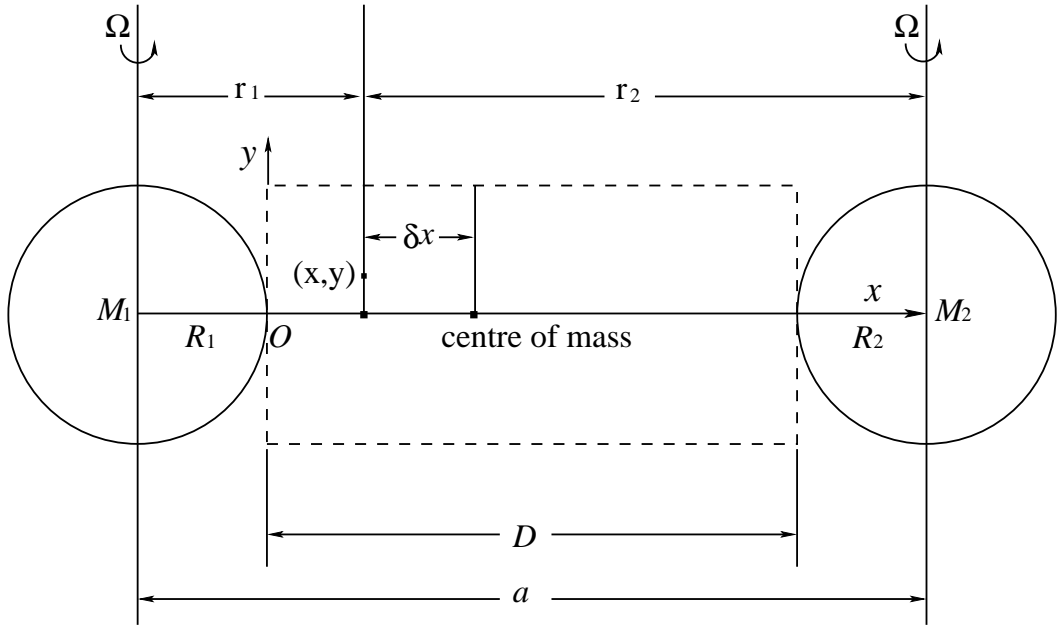}
  \caption{Sketch of the binary system. For the case of YY Gem,
  the two stars are identical. Here, $\delta
  \emph{x}$ is the deviation of the $x$-coordinate from the centre of
  mass, $r_{1}$ is the distance to the centre of $M_{1}$, and $r_{2}$ is
  the distance to the centre of $M_{2}$. The simulation box is indicated
  by the dashed box.
  }\label{fig1}
\end{figure*}

The basic model used in our work is shown in Fig. 1. For ordinary
binaries, the member stars are usually of different types. As
usual, we call the one with larger mass the primary, marked as
$M_1$, the other the secondary, marked as $M_2$ ($M_1 > M_2$).
Owing to the difference in mass, the point of zero gravity and the
centre of mass are located at different positions. The former is
closer to the secondary ($M_2$) and the latter is closer to the
primary ($M_1$). In the case of YY Gem, we take the parameters
given by \citet{Brancewicz80}, i.e., $M_1=M_2$, so that the two
points are actually cospatial. As shown in Fig. \ref{fig1}, $a$ is
the separation of two stellar centres, and $D=2.63$ R$_{\odot}$ is
the distance between the two surfaces along the joint line of the
stars.

Because of the short orbital period, \citet{Struve59} and
\citet{Qian02} suggested that YY Gem should be in synchronized
rotation. Therefore, we assume that the angular velocities of both
stars have the same value as that of the co-rotating reference
system. This means that the two stars rotate almost face to face
except the differential rotation. As a result of the differential
rotation, persistent shear motion is imposed to the coronal loops
whose footpoints are rooted at different latitudes.

The MHD processes driven by the differential rotation are
three-dimensional in nature. The addition of the two spherical
boundaries makes the problem more unfeasible to solve numerically.
To simplify the problem without losing the essence of physics, the
spherical boundaries are treated as two parallel planes, which are
tangent to the spheres at the equators as illustrated with the
dotted lines in Fig. \ref{fig1}. The area of interest is $0
\leqslant x \leqslant D$ and $-2.5$ R$_{\odot}\leqslant y
\leqslant 2.5$ R$_{\odot}$, where the $x$-axis is along the joint
line of the binary, the $y$-axis is parallel to the axis of the
rotation, and the $z$-axis is perpendicular to the $x-y$ plane,
with the  origin of the coordinate being located on the equator of
$M_1$ star. Note that the vertical size is set to be larger than
the stellar radius in order to minimize the influence of the top
boundary on the numerical results. Moreover, all quantities are
approximated to be invariant along the $z$-axis, so that the
problem becomes 2.5-dimensional. In reality, the interface between
the magnetic systems of the two stars has a limited extension in
the $z$-direction. Therefore, it should be kept in mind that any
feature in our numerical results should have a limited size in the
$z$-direction.

\subsection{MHD equations}

We perform 2.5-dimensional $(\partial/\partial z = 0)$ numerical
simulations in the Cartesian coordinates as described above.
The MHD equations are slightly modified from \citet{Chenpf99a,
Chenpf99b} in order to include the gravity from both stars and the
centrifugal force in the co-rotating system. For simplicity, heat
conduction is not included. The resulting dimensionless MHD
equations, which are shown below, are numerically solved by a
multistep implicit scheme \citep*{Hu89,chen00}.

\begin{equation}
    \frac{\partial \rho}{\partial t} + v_{x} \frac{\partial
    \rho}{\partial x} + v_{y} \frac{\partial \rho}{\partial
    y} + \rho \frac{\partial v_{x}}{\partial x} + \rho \frac{\partial
    v_{y}}{\partial y} = 0,
\end{equation}

\begin{equation}\label{m2}
    \frac{\partial v_{x}}{\partial t} + v_{x} \frac{\partial
    v_{x}}{\partial x} + v_{y} \frac{\partial v_{x}}{\partial
    y} + \frac{T}{\rho} \frac{\partial \rho}{\partial x} +
    \frac{\partial T}{\partial x} + \frac{2B_{z}}{\rho \beta_{0}} \frac{\partial B_{z}}
    {\partial x} + \frac{2}{\rho \beta_{0}} \frac{\partial \psi }
    {\partial x} \Delta \psi - g - F_{c} = 0,
\end{equation}

\begin{equation}\label{m3}
    \frac{\partial v_{y}}{\partial t} +v_{x} \frac{\partial v_{y}}
    {\partial x} + v_{y}\frac{\partial v_{y}}{\partial y} +
    \frac{\partial T}{\partial y} + \frac{T}{\rho} \frac{\partial
    \rho}{\partial y} + \frac{2}{\rho \beta_{0}} \frac{\partial
    \psi}{\partial y} \Delta \psi + \frac{2B_{z}}{\rho
    \beta_{0}} \frac{\partial B_{z}}{\partial y}= 0,
\end{equation}

\begin{equation}
    \frac{\partial v_{z}}{\partial t} + v_{x} \frac{\partial
    v_{z}}{\partial x} +v_{y} \frac{\partial v_{z}}{\partial y} +
    \frac{2}{\rho \beta_{0}} \frac{\partial \psi}{\partial x}
    \frac{\partial B_{z}}{\partial y} - \frac{2}{\rho \beta_{0}}
    \frac{\partial \psi}{\partial y} \frac{\partial B_{z}}{\partial
    x} = 0,
\end{equation}

\begin{equation}
    \frac{\partial \psi}{\partial t} + v_{x} \frac{\partial
    \psi}{\partial x} + v_{y} \frac{\partial \psi}{\partial y} -
    \frac{1}{R_{m}} \Delta \psi = 0,
\end{equation}

\begin{equation}
    \frac{\partial B_{z}}{\partial t} - \frac{\partial v_{z}}{\partial
    x} \frac{\partial \psi}{\partial y} + \frac{\partial v_{z}}
    {\partial y} \frac{\partial \psi}{\partial x} + B_{z}
    \frac{\partial v_{x}}{\partial x} + B_{z} \frac{\partial
    v_{y}}{\partial y} + v_{x} \frac{\partial B_{z}}{\partial x} +
    v_{y} \frac{\partial B_{z}}{\partial y} - \frac{\partial}{\partial
    x} \left(\frac{1}{R_{m}} \frac{\partial B_{z}}{\partial x}\right) -
    \frac{\partial}{\partial y} \left(\frac{1}{R_{m}} \frac{\partial B_{z}}{\partial
    y}\right) = 0,
\end{equation}

\begin{equation}
    \frac{\partial T}{\partial t} + v_{x} \frac{\partial T}{\partial
    x} + v_{y} \frac{\partial T}{\partial y} + (\gamma - 1)T
    \frac{\partial v_{x}}{\partial x} + (\gamma -1)T \frac{\partial
    v_{y}}{\partial y} - \frac{2(\gamma - 1)}{\rho \beta_{0} R_{m}}
     \left[\left(\frac{\partial B_{z}}{\partial
    x}\right)^{2} + \left(\frac{\partial B_{z}}{\partial y}\right)^{2} + (\Delta
    \psi)^2\right] = 0,
\end{equation}
\noindent where $\rho$, $v_{x}$, $v_{y}$, $v_{z}$, $\psi$, $B_{z}$
and $T$ correspond to the dimensionless density, three components
of velocity, magnetic flux function, $z$-component of the magnetic
strength, and temperature, respectively, and $\rho_{0}$, $v_{0}$,
$\psi_{0}$, $B_{0}$ and $T_{0}$ are the characteristic values of
the corresponding parameters, which are used to nondimensionalize
the equations. The characteristic velocity $v_{0}$ is defined as
the isothermal sound speed $v_{0} = \sqrt{RT_0}$, where $R$ is the
gas constant for the fully ionised hydrogen and the length scale
$L_{0}$ is set to be 0.5 R$_{\odot}$. The dimensionless size of
the simulation box is then $0\leq |y| \leq 5$ and $0\leq x \leq
5.26$. Since the flux function and the magnetic field are related
by
\begin{equation}
    \textbf{\emph{B}} = \nabla \times (\psi \textbf{\emph{e}}_{z})
    + B_{z} \textbf{\emph{e}}_{z},
\end{equation}
where $\textbf{\emph{e}}_{z}$ is the unit vector in the
$\emph{z}$-axis, we have $\psi_{0} = B_{0} L_{0}$.

In equation (\ref{m2}), $g$ is the total acceleration of gravity
contributed by both stars,
\begin{equation}
    g = \frac{G }{v_{0}^{2} L_{0}} (\frac{M_{2}}{r_{2}^{2}} - \frac{M_{1}}{r_{1}^{2}}),
\end{equation}
where $M_{1}$ and $M_{2}$ are the masses of the two stars, $r_1$
and $r_2$ are indicated in Fig. \ref{fig1} (note that such an
expression is an extension of the gravity along the $x$-axis,
which is valid since we are interested in the processes near the
$x$-axis). In the same equation, $F_{c}$ represents the
centrifugal force,
\begin{equation}
    F_{c} = \frac{G (M_{1} + M_{2}) L_{0}^{2}}{v_{0}^{2}
    a^3} \delta x,
\end{equation}
where $a$ is the distance between the two stars, $\delta \emph{x}$
is the deviation of the $x$-coordinate from the centre of mass.
The plasma beta, $\beta_{0}$, and the magnetic Reynolds number,
$R_m$, are expressed as
\begin{equation}
    \beta_{0} = \frac{2\mu_{0}\rho_{0}v_{0}^{2}}{B_{0}^{2}},
\end{equation}
\begin{equation}
    R_{m} = \frac{\mu_{0}v_{0}L_{0}}{\eta},
\end{equation}
where $\mu_{0}$ is the magnetic permeability and $\eta$ is the
anomalous resistivity. Similarly to \citet{Chenpf06}, $\eta$ is
chosen to be a function of the current density, $j$,
\begin{equation}
    \eta = \left\{
\begin{array}{cc}
  \eta_{0} \min(1, j/j_{c} - 1), & j \geqslant j_{c}, \\
  0, & j < j_{c}, \\
\end{array}\right.
\end{equation}
where $j_{c}$ is the critical current density.

The time unit used in the description of the numerical results is the
Alfv\'{e}n time-scale, $\tau_{A} = L_{0}/v_{A}$, where $v_{A}$ is
Alfv\'{e}n speed, i.e.,
\begin{equation}
    v_A = \frac{\psi_{0}}{L_{0}\sqrt{\mu_{0}\rho_{0}}}.
\end{equation}

As \citet{Tsikoudi00} pointed out, the corona of YY Gem is very
hot, $T_0$ can be taken to be several million kelvin. The number
density ($n=\rho/m_p$) of the stellar corona is in the range of
$10^{14}$ -- $10^{18}$ m$^{-3}$ (\citealt{Monsignori94};
\citealt{Mewe95}; \citealt{Schmitt96}; \citealt{Schrijver00};
\citealt{Ness02}). The typical strength of the general magnetic
field near the stellar surface is tens of Gauss. Therefore, in our
standard case, which is called case A2 later, the following
characteristic values are adopted: $n_0=10^{15}$ m$^{-3}$,
$T_0=1.9\times 10^6$ K, $\beta_0$=0.01 (hence $B_0$=36 G), and
$j_c=10$. The corresponding time-scale is $\tau_A=139$ s.

\subsection{Initial and boundary conditions}

\begin{figure}
  \centering
  \includegraphics[width=9cm,height=5cm]{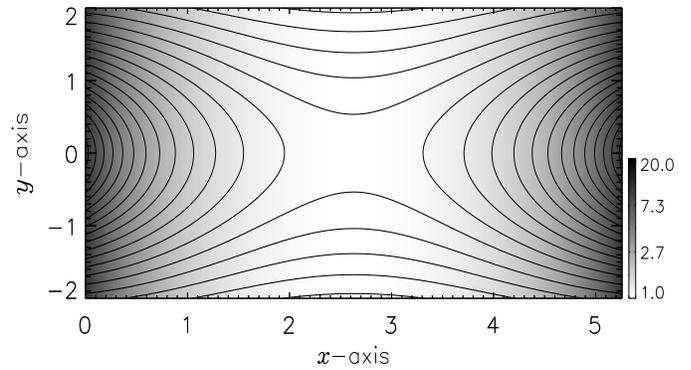}\\
  \caption{Initial configuration of the magnetic field (solid lines) and
  the distribution of the dimensionless density (gray scale).}\label{fig2}
\end{figure}

Initially, the isothermal plasma is at rest, i.e.,
$v_x=v_y=v_z=0$, and $T=1$. The density distribution is obtained
by numerically calculating the hydrostatic equation
\begin{equation}
    \frac{1}{\rho} \frac{\textrm{d}\rho}{\textrm{d}x} = g(x) + F_{c}(x).
\end{equation}

The initial magnetic field in this work is assumed to be
potential, which is produced by two parallel line currents lying
below the two surfaces, i.e.,

\begin{equation}\label{mag}
    \psi =-\ln(\sqrt{(x - x_{1})^2 + y^2}) - \ln(\sqrt{(x -
    x_{2})^2 + y^2}),
\end{equation}
where $x_1$ and $x_2$ are the $x$-coordinates of these two
currents. We take $x_1 = -1$ and $x_2 = D/L_0 + 1$ in our work.
The resulting density distribution and the magnetic configuration
are plotted in Fig. \ref{fig2}, where the magnetic field is
similar to the configuration adopted in \citet{Ferreira98}. The
magnetic configuration is taken in such a way that a magnetic null
point exists between the antiparallel magnetic loops. So, whenever
the closed loops bulge, a current sheet forms natually, and
magnetic reconnection is ready to commence (if the polarity of one
magnetic system expressed in equation (\ref{mag}) is flipped, the
two loop systems become to repel each other, and no reconnection
can happen). Similar magnetic connectivity with a null point was
also computed for the active RS CVn system (\citealt{Uchida83};
\citealt{Uchida85}; \citealt{Beasley00}).

Owing to the symmetry, calculations are done only in the upper
half of the simulation area, i.e., $0 \leqslant x \leqslant D$ and
$0\leqslant y \leqslant 2.5$ R$_{\odot}$. The calculation domain
is discretised into 501 uniform grid points along the $x$-axis and
141 nonuniform grid points along the $y$-axis, with more grid
points concentrated near the $x$-axis.  The top side of the
simulation box ($y=5$) is treated as an open boundary, and the
bottom one ($y=0$) is a symmetry boundary. The right and left,
which correspond to the surface of the binary stars, are line-tied
boundaries, where all quantities are fixed except that a typical
differential rotation is imposed at each surface in the following
form:

\begin{equation}
    v_z|_{\textrm{left}} = \left\{
    \begin{array}{cc}
      - \Omega R(1-\alpha \sin^{2}i) \sqrt{1-\sin^{2}i} + \Omega R ,& i \leqslant \pi/2,\\
      \Omega R,& i > \pi/2, \\
    \end{array}\right.
\end{equation}

\begin{equation}
    v_z|_{\textrm{right}} = \left\{
    \begin{array}{cc}
      \Omega R(1-\alpha \sin^{2}i)\sqrt{1-\sin^{2}i} - \Omega R ,& i \leqslant \pi/2,\\
      - \Omega R,& i > \pi/2, \\
      \end{array}\right.
\end{equation}
where $\Omega$ is the angular velocity of the binary, $\alpha$ is
the differential rotation rate, and $i$, the latitude, is related
to the coordinate $y$ by $y=R\sin i$. Beyond $y=R$, $v_z$ is set
to be the value at $y=R$. In this paper, $\alpha$ is set to be
0.5. As we will discuss later, the value of $\alpha$ does not
affect the main results in our numerical simulations.

\section{numerical results}

\subsection{General evolution}

\begin{figure*}
  \centering
  \includegraphics[width=6cm,height=15cm]{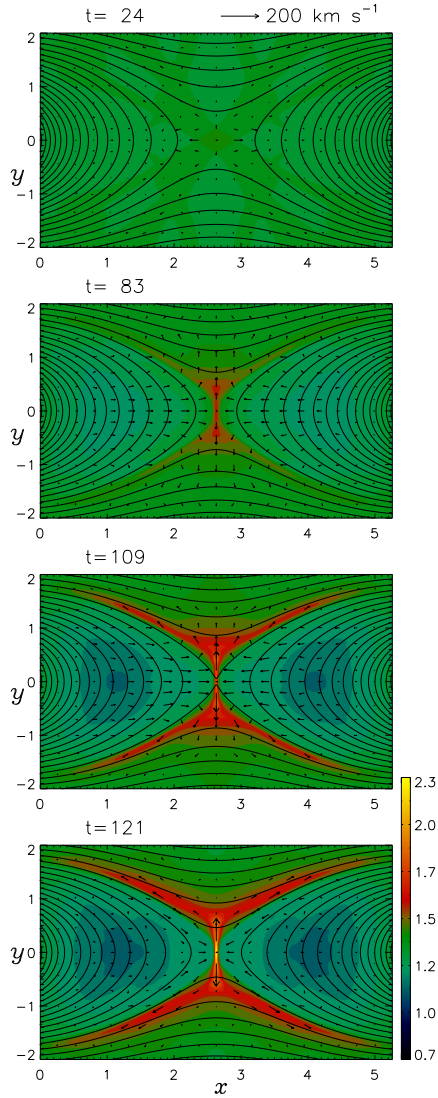}\\
  \caption{Magnetic field lines (solid lines), velocities (arrows),
  and temperatures (colour) at five selected times in case A2.}
\label{fig3}
\end{figure*}

\begin{figure}
  \centering
  \includegraphics[width=8cm]{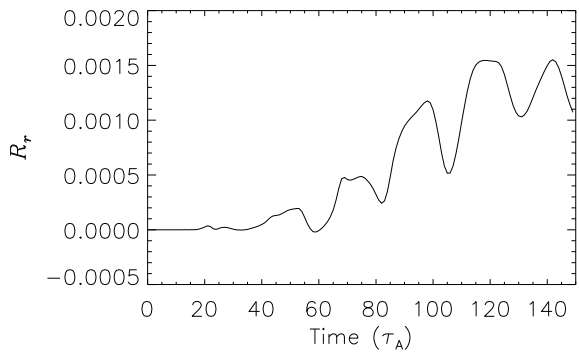}\\
  \caption{Temporal evolution of the reconnection rate in case A2, where
  $R_r$ is the reconnection rate, and $\tau_A=139$ s.}\label{A2rate}
\end{figure}

The distributions of the magnetic field, the temperature, and the
velocity field at several selected times in case A2 are shown in
Fig. \ref{fig3}. As the shear motion is imposed at the left and
right boundaries, footpoints of the coronal loops are dragged to
move in the $z$-direction, which increases the local magnetic
pressure. As a result, both the left and the right magnetic loop
systems begin to inflate and approach each other due to the
increasing pressures near the roots of the loop systems (see the
snapshot at $t=83$ $\tau_A$). Such an inflation of magnetic loops
due to shear motions was demonstrated on the Sun \citep{Barnes72}.
The two loop systems collide near the magnetic null point, by
which a current sheet forms. When the increasing current density
exceeds the prescribed threshold ($j_c$), anomalous resistivity is
excited, which triggers the occurrence of magnetic reconnection
between the two loop systems (see the snapshot at $t=109$
$\tau_A$). The temporal evolution of the reconnection rate, $R_r$,
which is defined as $\textrm{d}\psi/\textrm{d}t$ at the magnetic
null point, is shown in Fig. \ref{A2rate}. It is seen that the
reconnection rate presents a periodic behaviour. With the wavelet
analysis, the period is found to be $54 \pm 6$ min. Note that,
with the 2.5D approximation in our model, the reconnection between
the two closed magnetic systems, which are rooted on the surfaces
of the binary stars, will go on as long as the differential
rotation continues. In reality, the differential rotation of the
two spherical stellar surfaces would take the two closed magnetic
systems apart, which would cease the reconnection after a certain
time. The 2.5D model is valid here since we are interested in the
magnetic reconnection only during first several periods.

In order to compare the numerical result with the \emph{U}-band
observations made by \citet{Doyle90a}, we synthesised the
\emph{UV} emission in the wave-band of $3650\pm10${\AA} from our
numerical results with the help of the CHIANTI database
(\citealt{dere97}; \citealt{land06}; see \citealt{Chenpf06} for
details). The calculated \emph{UV} light-curve of the whole
simulation area is plotted in the left panel of Fig. 5, and its
wavelet spectrum is presented in the right panel. Note that since
the reconnection rate and the \emph{UV} emission both are small in
the first 60 $\tau_A$, we displayed the result after $t=60
\tau_A$. It is seen that the \emph{UV} flux does show a
periodicity, and the period is centred at 52 min, very close to
that of the reconnection rate, i.e., 54 min.

\begin{figure*}
  \centering
  \includegraphics[width=14cm]{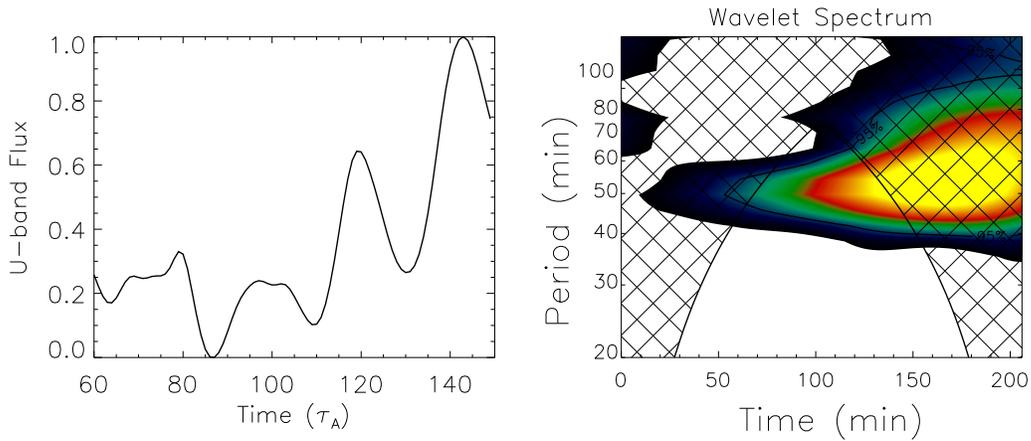}\\
  \caption{Left panel: Temporal evolution of the flux in the \emph{U}-band
  , which is derived by using the CHIANTI database. Right
  panel: The wavelet spectrum corresponding to the \emph{U}-band
  flux, which has a period centred around about 52 min.
  }\label{urad}
\end{figure*}

\subsection{Parameter survey}

In order to investigate how the choice of the characteristic
values of various parameters may affect the results, we have made
numerical simulations with different characteristic values, which
are listed in Table \ref{tab1}. In cases A1--A5, only $T_0$ is
changed. Similarly, cases B1--B5 are for $\beta_0$ dependence,
cases C1--C3 for $\rho_0$, and cases D1--D3 for $j_c$,
respectively. In all cases, $L_0$ is fixed to be 0.5 R$_{\odot}$.
Numerical simulations indicate that, similar to case A2 as
described in the previous subsection, all other cases present
periodic behaviours. The corresponding period of each case is
displayed in the last column of Table \ref{tab1}. It is found
that, with $T_0$, $\beta_0$, $\rho_0$, and $j_c$ as independent
parameters, the period of the flaring is strongly dependent on
$T_0$ and $\beta_0$, and an increasing $T_0$ or decreasing
$\beta_0$ results in a decreasing period. When $\rho_0$ increases,
the magnetic field $B_0$ increases as well in order to keep
 $\beta_0$ constant (correspondingly, the Alfv\'en speed is constant).
As a result, the period of the reconnection rate does not
change.It is also seen from Table \ref{tab1} that $j_c$ has little
effect on the period, which implies that the magnetic reconnection
is modulated by some external process.

\begin{table}
    \centering
    \caption{Parameters used in different simulation cases.}
    \begin{tabular}{|c|c|c|c|c|c|}
    \hline
   Models & $T_{0}$ ($\times 10^{6}$ K) & $\beta_{0}$ & $\lg n_{0}$/m$^{-3}$ & $j_{c}$  & $P$ (min)  \\
  \hline
  A1 & 1.5 & 0.01 & 15 & 10.0 & 70 $\pm$ 9 \\
  A2 & 1.9 & 0.01 & 15 & 10.0 & 54 $\pm$ 6 \\
  A3 & 2.5 & 0.01 & 15 & 10.0 & 41 $\pm$ 4 \\
  A4 & 3.0 & 0.01 & 15 & 10.0 & 35 $\pm$ 3 \\
  A5 & 4.0 & 0.01 & 15 & 10.0 & 31 $\pm$ 3 \\
  \hline
  B1 & 2.0 & 0.05 & 15 & 10.0 & 96 $\pm$ 15\\
  B2 & 2.0 & 0.025 & 15 & 10.0 & 78 $\pm$ 10\\
  B3 & 2.0 & 0.01 & 15 & 10.0 & 52 $\pm$ 6 \\
  B4 & 2.0 & 0.0075 & 15 & 10.0 & 47 $\pm$ 4 \\
  B5 & 2.0 & 0.005 & 15 & 10.0 & 38 $\pm$ 4 \\
  \hline
  C1 & 2.0 & 0.01 & 14 & 10.0 & 52 $\pm$ 6 \\
  C2 & 2.0 & 0.01 & 15 & 10.0 & 52 $\pm$ 6 \\
  C3 & 2.0 & 0.01 & 16 & 10.0 & 52 $\pm$ 6 \\
  \hline
  D1 & 2.0 & 0.01 & 15 & 1.0 & 52 $\pm$ 6 \\
  D2 & 2.0 & 0.01 & 15 & 5.0 & 52 $\pm$ 6 \\
  D3 & 2.0 & 0.01 & 15 & 10.0 & 52 $\pm$ 6 \\
  \hline
\end{tabular}\label{tab1}
\end{table}

Besides the four parameters mentioned above, we also examine other
parameters related to the binary itself. In case A2, by changing
the values of $\alpha$ ($\alpha = 0.2$, $0.3$, $0.4$, $0.5$,
$0.6$, $0.7$, $0.8$), we find that the results keep invariant,
i.e., $P=54 \pm 6$ min. In other words, the flaring period has
nothing to do with the speed of the differential rotation. As
mentioned above, the mass and the size of YY Gem are taken from
\citet{Brancewicz80}. However, other authors derived slightly
different values (\citealt{Leung78}; \citealt{Chabrier95};
\citealt{Torres02}), e.g., in \citet{Leung78}, $M_{1} =0.62$
M$_{\odot}$, $M_{2} = 0.57$ M$_{\odot}$, $R_{1} = 0.66$
R$_{\odot}$, $R_{2} = 0.58$ R$_{\odot}$, the separation of two
components $a = 3.9$ R$_{\odot}$, and the orbital period $P =
0.82$ d. Through test calculations, we find little change in the
results.

\section{Discussions}

\subsection{What modulates the magnetic reconnection?}
\begin{figure}
  \centering
  \includegraphics[width=8cm,height=6cm]{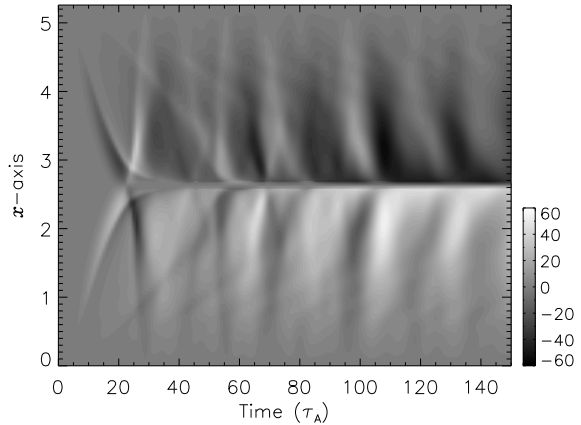}\\
  \caption{Temporal evolution of $v_{x}$ along the $x$-axis.}\label{vxgray}
\end{figure}

In order to understand how the magnetic reconnection is modulated,
the temporal evolution of $v_{x}$ along the $x$-axis is shown in
Fig. \ref{vxgray}, where the value of $v_x$ is indicated by the
grey scales. As the reconnection rate $R_r$ is equivalent to
$v_xB_y$ along the $x$-axis, it is expected to see that $v_x$ at a
given point also shows an oscillatory evolution, with the same
period as $R_r$, i.e., 54 min. Furthermore, a wave pattern is
discerned in Fig. \ref{vxgray}, where the waves are seen to travel
back and forth in the $x$-direction, bounded by the surfaces of
the two stars. Since the waves propagate perpendicular to the
magnetic field lines, we postulate that they are fast-mode
magnetoacoustic waves, and that the periodic magnetic reconnection
in the numerical results is modulated by such waves.

\begin{figure}
  \centering
  \includegraphics[width=8cm]{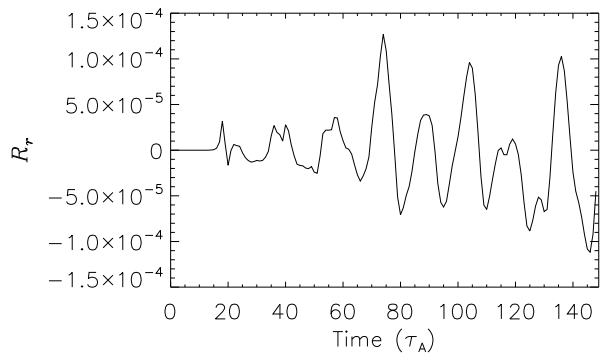}\\
  \caption{Temporal evolution of reconnection rate in the case that
  the horizontal size of the simulation box is cut down.
  }\label{d_cut}
\end{figure}

To verify that the magnetic reconnection is modulated by
magnetoacoustic waves bouncing back and forth between the two
stellar surfaces, we shift the right boundary of case A2 leftward
and the left boundary rightward by 0.8 simultaneously, by which
the separation between the two stellar surfaces is cut down from
5.26 to 3.66, i.e., a decrease with a ratio of 30 per cent. The
corresponding evolution of the reconnection rate is plotted in
Fig. \ref{d_cut}. Note that the positive $R_r$ means that the
field lines in the two loop systems are reconnecting, whereas the
negative $R_r$ means that the transverse field lines are
reconnecting.

It is found that the resulting period of magnetic reconnection
becomes 34 $\pm$ 6 min. This means that the period decreases by 37
per cent, a ratio close to that of the decrease of the separation
between the two stellar surfaces. This implies that the magnetic
reconnection is modulated by waves trapped between the two stellar
surfaces.

To verify that the travelling waves are fast-mode waves, here we
calculate the trajectory of fast-mode waves propagating from the
reconnection point ($x=2.63$), through the right boundary
($x=5.26$) and then back to the reconnection point along the
$x$-axis. For a given number $n$, the travelling time ($T$) and
the position of the wave front ($x$) are determined by the
following equations

\begin{equation}\label{tw1}
        T =\sum_{i\geqslant 1}^{n}\Delta x/v_f, \\
        x-2.63=\sum_{i\geqslant 1}^{n}\Delta x,
\end{equation}
\noindent where $\Delta x$ is the uniform grid spacing in the
$x$-direction, and $v_f$ is the fast-mode wave velocity. The
procedure is described as follows: Starting from the reconnection
point ($x$=2.63), the wave front propagates rightward by $\Delta
x$ each step, the time interval $\Delta t$ is equal to $\Delta x$
divided by $v_f$, so the total travelling time is the sum of
$\Delta t$, while the position is the sum of $\Delta x$. Note that
since both the magnetic field and the plasma temperature are
changing with time, we update the magnetic and thermal parameters
every 1 $\tau_A$ when calculating $v_f$. Besides, when the wave
reaches the right boundary and then bounces back, $\Delta x$ takes
a negative value. After the wave front returns to the reconnection
point, the wave should propagate to the left half area of the
simulation box. Considering the symmetry of the problem, i.e.,
there is a symmetric wave propagating rightward from the left half
area, we repeat the above calculation in the right half region.

\begin{figure}
  \centering
  \includegraphics[width=8cm]{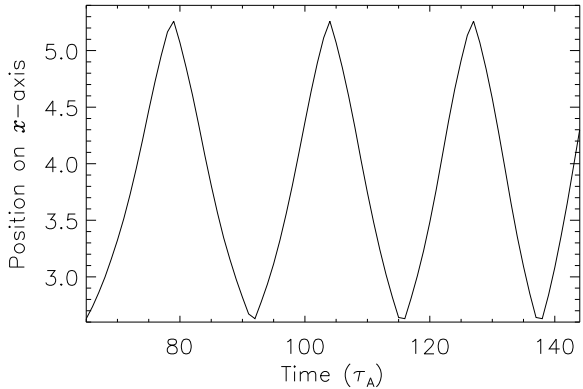}\\
  \caption{Temporal evolution of the wave front. Only the right half is
  plotted owing to the symmetry of the YY Gem system.
  }\label{front}
\end{figure}

Figure. \ref{front} shows the temporal evolution of the position
of the wave front in the right half region only. It is found that
the trajectory of the fast-mode wave, from the reconnection point
to the right (or left) boundary and then back to the reconnection
point, is quasi-periodic, with a period of $\sim23\pm2$ $\tau_A$,
i.e., $\sim 53 \pm 5$ min, very close to the period of the
magnetic reconnection rate presented in Section 3. Therefore, we
are convinced that the magnetic reconnection is modulated by
fast-mode waves. The underlying process can be understood as
follows: When symmetric waves, introduced either by the shear
motions of the stellar surfaces or by the magnetic reconnection,
collide near the reconnection point, the reconnection rate is
enhanced to form a peak. After the collision, the two waves
propagate to the right and left boundaries and are bounced back,
respectively. When they collide again near the reconnection point,
the reconnection rate is enhanced again. Therefore, the period of
the magnetic reconnection is determined by the travelling time of
fast-mode waves from the reconnection point to the right (or left)
boundary and then back to the reconnection point.

\subsection{Empirical formula for the period}

As shown in Section 3, among the selected four parameters, the
period of the reconnection rate ($P$) is explicitly dependent on
$T_0$ and $\beta_0$. The underlying process can be understood as
follows: When $T_0$ increases while keeping $\beta_0$ and $\rho_0$
constant, both the sound speed and the Alfv\'en speed increase,
which leads to a larger fast-mode wave speed. As a result, the
period of the reconnection rate decreases; When $\beta_0$
decreases while keeping $T_0$ and $\rho_0$ constant, the sound
speed does not change, while the Alfv\'en speed increases, which
also leads to a larger fast-mode wave speed and a shorter period;
When $\rho_0$ increases while keeping $\beta_0$ and $T_0$
constant, both the sound and the Alfv\'en speed do not change (the
Alfv\'en speed does not change since the magnetic field $B_0$
increases). As a result, the period keeps constant.

\begin{figure}
  \centering
  \includegraphics[width=8cm]{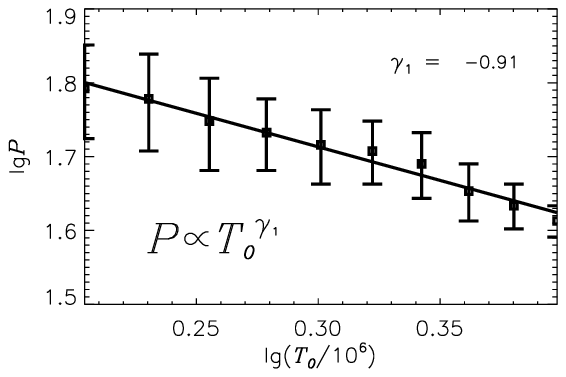}\\
  \includegraphics[width=8cm]{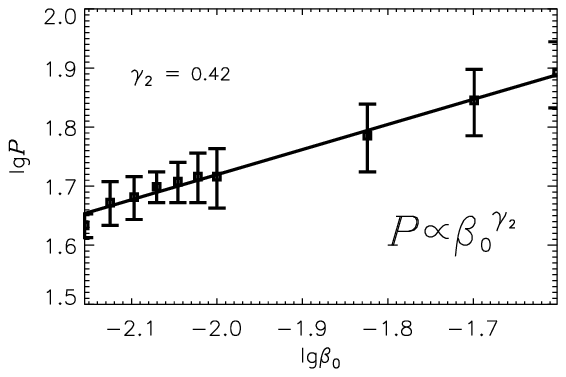}\\
  \caption{Upper panel: Relation between the period ($P$) and the
           temperature ($T$); Lower panel: Relation between the period
       ($P$) and $\beta_0$. The solid lines are the corresponding
       linear fitting of the data points.
  }\label{P-T}
\end{figure}

In order to derive an empirical formula to relate $P$ to $T_0$ and
$\beta_0$, we simulate a series of cases with different $T_0$ and
$\beta_0$. The corresponding variations of $P$ with $T_0$ and
$\beta_0$ are displayed by the squares in Fig. \ref{P-T}. It is
seen that $\lg P$ is almost a linear function of $\lg T_0$ and
$\lg \beta_0$, which means that $P$ scales with $T_0$ and
$\beta_0$ as $P \propto T_0^{\gamma_1} \beta_0^{\gamma_2}$. By
fitting the data points with the solid lines in Fig. \ref{P-T}, we
obtain $\gamma_1=-0.91$ and $\gamma_2=0.42$. Therefore, we have $P
\propto T_0 ^{-0.91}\beta_0^{0.42}$, i.e.,

\begin{equation}\label{emfor}
    P \sim \rho_0^{0.42} T_0^{-0.49} B_0^{-0.84}.
\end{equation}

According to our postulation that the magnetic reconnection is
modulated by fast-mode waves, the period $P$ should be
proportional to $1/v_f$, where $v_f^2=c_s^2+v_A^2$, $c_s^2=\gamma
RT_0$, and $v_A^2=B_0^2/(\mu_0 \rho_0)$. In the extreme case of
zero magnetic field, we have $P\propto T_0^{-0.5}$; In the case of
extremely strong magnetic field, we have $P\propto
\rho_0^{0.5}B_0^{-1}$. The slight deviation of the power indices
in our numerical results from those in the two extreme cases
further implies that it is the fast-mode waves that modulate the
magnetic reconnection. Based on this scaling law, the observed
period of the flaring rate in YY Gem, i.e., 48 min, corresponds to
a characteristic magnetic field of 41 G, which is the value of the
magnetic field in low corona, providing the number density of the
plasma in the inter-binary space is about $n_0 \sim 10^{15}$
m$^{-3}$.

\subsection{Other points}

The periodic flaring of YY Gem, which was found by
\citet{Doyle90a}, was not confirmed in \citet{Stelzer02}. The
discrepancy might be attributed to two factors: (1) Probably the
two observations were made at different phases of the stellar
magnetic cycle. In order to have the large-scale magnetic
reconnection in the inter-binary space, the magnetic configuration
of each star should be close to the bipolar type, with
interconnecting magnetic field lines linking the polar regions of
the binary stars, as shown by Fig. \ref{fig2}. Such a magnetic
configuration often appears near activity minimum; (2) The
periodic flares observed by \citet{Doyle90a} and confirmed by our
numerical simulations, are produced in the inter-binary space,
therefore, are large-scale in nature. Without doubt, there should
be small-scale flares that occur near the stellar surface and are
due to the reconnection of the magnetic fields of an individual
star. The flares observed by \citet{Stelzer02} probably correspond
to these small-scale flares.

\section{Conclusions}

Observations indicated that there may exist periodicity in the
occurrence of strong flares from YY Gem. Using 2.5-dimensional MHD
numerical simulations, we confirmed the periodicity of the flaring
in the inter-binary space of YY Gem. On the basis of the analysis
of the simulation results, the following conclusions can be drawn:

1. Owing to the differential rotation of the stellar surface, coronal
   loops in each star inflate and approach each other. The resulting
   magnetic reconnection generates the inter-binary flares, which are
   large-scale in nature. Fast magnetoacoustic waves, which are trapped
   in the inter-binary space between the surfaces of the two binary
   components, modulate the magnetic reconnection, producing the
   periodic behaviour of the flaring rate.

2. The period scales with the typical coronal density ($\rho_0$),
   temperature ($T_{0}$), and the coronal magnetic field ($B_0$)
   as $P\sim \rho_0^{0.42} T_0^{-0.49} B_0^{-0.84}$.

\section*{Acknowledgments}
We thank the referee for constructive comments that led to an
improvement of the paper. We are also grateful to C. Fang, X. Y.
Xu, M. Jin, and Q. M. Zhang for their helpful suggestions. This
work was supported by the Chinese foundations NCET-04-0445, FANEDD
(200226), NSFC under grants 10025315, 10221001, 10333040,
10403003, and 10673004, and by NKBRSF under grant 2006CB806302.


\bsp

\label{lastpage}

\end{document}